\documentclass[12pt]{article}
\usepackage{epsfig}
\usepackage{graphicx}
\usepackage{amsmath}
\usepackage{hhline}
\usepackage{amssymb}
\usepackage{times}
\usepackage{cite}

\newlength{\dinwidth}
\newlength{\dinmargin}
\begin{document}  
\newcommand{\pom}{{I\!\!P}}
\newcommand{\reg}{{I\!\!R}}
\newcommand{\slowpi}{\pi_{\mathit{slow}}}
\newcommand{\fiidiii}{F_2^{D(3)}}
\newcommand{\fiidiiiarg}{\fiidiii\,(\beta,\,Q^2,\,x)}
\newcommand{\n}{1.19\pm 0.06 (stat.) \pm0.07 (syst.)}
\newcommand{\nz}{1.30\pm 0.08 (stat.)^{+0.08}_{-0.14} (syst.)}
\newcommand{\fiidiiiful}{F_2^{D(4)}\,(\beta,\,Q^2,\,x,\,t)}
\newcommand{\fiipom}{\tilde F_2^D}
\newcommand{\ALPHA}{1.10\pm0.03 (stat.) \pm0.04 (syst.)}
\newcommand{\ALPHAZ}{1.15\pm0.04 (stat.)^{+0.04}_{-0.07} (syst.)}
\newcommand{\fiipomarg}{\fiipom\,(\beta,\,Q^2)}
\newcommand{\pomflux}{f_{\pom / p}}
\newcommand{\nxpom}{1.19\pm 0.06 (stat.) \pm0.07 (syst.)}
\newcommand {\gapprox}
   {\raisebox{-0.7ex}{$\stackrel {\textstyle>}{\sim}$}}
\newcommand {\lapprox}
   {\raisebox{-0.7ex}{$\stackrel {\textstyle<}{\sim}$}}
\def\gsim{\,\lower.25ex\hbox{$\scriptstyle\sim$}\kern-1.30ex%
\raise 0.55ex\hbox{$\scriptstyle >$}\,}
\def\lsim{\,\lower.25ex\hbox{$\scriptstyle\sim$}\kern-1.30ex%
\raise 0.55ex\hbox{$\scriptstyle <$}\,}
\newcommand{\pomfluxarg}{f_{\pom / p}\,(x_\pom)}
\newcommand{\dsf}{\mbox{$F_2^{D(3)}$}}
\newcommand{\dsfva}{\mbox{$F_2^{D(3)}(\beta,Q^2,x_{I\!\!P})$}}
\newcommand{\dsfvb}{\mbox{$F_2^{D(3)}(\beta,Q^2,x)$}}
\newcommand{\dsfpom}{$F_2^{I\!\!P}$}
\newcommand{\gap}{\stackrel{>}{\sim}}
\newcommand{\lap}{\stackrel{<}{\sim}}
\newcommand{\fem}{$F_2^{em}$}
\newcommand{\tsnmp}{$\tilde{\sigma}_{NC}(e^{\mp})$}
\newcommand{\tsnm}{$\tilde{\sigma}_{NC}(e^-)$}
\newcommand{\tsnp}{$\tilde{\sigma}_{NC}(e^+)$}
\newcommand{\st}{$\star$}
\newcommand{\sst}{$\star \star$}
\newcommand{\ssst}{$\star \star \star$}
\newcommand{\sssst}{$\star \star \star \star$}
\newcommand{\tw}{\theta_W}
\newcommand{\sw}{\sin{\theta_W}}
\newcommand{\cw}{\cos{\theta_W}}
\newcommand{\sww}{\sin^2{\theta_W}}
\newcommand{\cww}{\cos^2{\theta_W}}
\newcommand{\trm}{m_{\perp}}
\newcommand{\trp}{p_{\perp}}
\newcommand{\trmm}{m_{\perp}^2}
\newcommand{\trpp}{p_{\perp}^2}
\newcommand{\alp}{\alpha_s}

\newcommand{\alps}{\alpha_s}
\newcommand{\sqrts}{$\sqrt{s}$}
\newcommand{\LO}{$O(\alpha_s^0)$}
\newcommand{\Oa}{$O(\alpha_s)$}
\newcommand{\Oaa}{$O(\alpha_s^2)$}
\newcommand{\PT}{p_{\perp}}
\newcommand{\JPSI}{J/\psi}
\newcommand{\sh}{\hat{s}}
\newcommand{\uh}{\hat{u}}
\newcommand{\MP}{m_{J/\psi}}
\newcommand{\PO}{I\!\!P}
\newcommand{\xbj}{x}
\newcommand{\xpom}{x_{\PO}}
\newcommand{\ttbs}{\char'134}
\newcommand{\xpomlo}{3\times10^{-4}}  
\newcommand{\xpomup}{0.05}  
\newcommand{\dgr}{^\circ}
\newcommand{\pbarnt}{\,\mbox{{\rm pb$^{-1}$}}}
\newcommand{\gev}{\,\mbox{GeV}}
\newcommand{\WBoson}{\mbox{$W$}}
\newcommand{\fbarn}{\,\mbox{{\rm fb}}}
\newcommand{\fbarnt}{\,\mbox{{\rm fb$^{-1}$}}}
%
%
\newcommand{\qsq}{\ensuremath{Q^2} }
\newcommand{\gevsq}{\ensuremath{\mathrm{GeV}^2} }
\newcommand{\et}{\ensuremath{E_t^*} }
\newcommand{\rap}{\ensuremath{\eta^*} }
\newcommand{\gp}{\ensuremath{\gamma^*}p }
\newcommand{\dsiget}{\ensuremath{{\rm d}\sigma_{ep}/{\rm d}E_t^*} }
\newcommand{\dsigrap}{\ensuremath{{\rm d}\sigma_{ep}/{\rm d}\eta^*} }
\newcommand{\dedx}{\ensuremath{{\rm d} E/{\rm d} x}}
\def\Journal#1#2#3#4{{#1} {\bf #2} (#3) #4}
\def\NCA{Nuovo Cimento}
\def\RPP{Rep. Prog. Phys.}
\def\ARNPS{Ann. Rev. Nucl. Part. Sci.}
\def\NIM{Nucl. Instrum. Methods}
\def\NIMA{{Nucl. Instrum. Methods} {\bf A}}
\def\NPB{{Nucl. Phys.}   {\bf B}}
\def\NPPS{Nucl. Phys. Proc. Suppl.} 
\def\NPPSC{{Nucl. Phys. Proc. Suppl.} {\bf C}}
\def\PR{Phys. Rev.}
\def\PLB{{Phys. Lett.}   {\bf B}}
\def\PRL{Phys. Rev. Lett.}
\def\PRD{{Phys. Rev.}    {\bf D}}
\def\PRC{{Phys. Rev.}    {\bf C}}
\def\ZPC{{Z. Phys.}      {\bf C}}
\def\EJC{{Eur. Phys. J.} {\bf C}}
\def\EPL{{Eur. Phys. Lett.} {\bf}}
\def\CPC{Comp. Phys. Commun.}
\def\NP{{Nucl. Phys.}}
\def\JPG{{J. Phys.} {\bf G}} 
\def\EPC{{Eur. Phys. J.} {\bf C}}
\def\PRSL{{Proc. Roy. Soc.}} {\bf}
\def\PETF{{Pi'sma. Eksp. Teor. Fiz.}} {\bf}
\def\JETPL{{JETP Lett}}{\bf}
\def\IJTP{Int. J. Theor. Phys.}
\def\HJ{Hadronic J.}

\noindent
Title of paper ``COSMIC RAYS AND CLIMATE CHANGE OVER THE PAST 1000 MILLION YEARS'' \\
\\
by T. Sloan (corresponding author)\\
    Dept of Physics, \\
    Lancaster University,\\
    Lancaster LA1 4YB, \\
    UK. \\
email t.sloan@lancaster.ac.uk \\
Telephone number (44)(0)17687 74467 \\
Fax number (44)(0)1524 844037 \\
\\
Second author, \\
A.W. Wolfendale, \\
Dept of Physics, \\
Durham University,\\ 
Rochester Building,\\ 
South Road, \\
Durham DH1 3LE,\\
UK.\\
email a.w.wolfendale@durham.ac.uk \\

\newpage

 


\begin{flushleft}
\end{flushleft}
\begin{center}
\begin{Large}
{\boldmath \bf COSMIC RAYS AND CLIMATE CHANGE OVER THE PAST 1000 MILLION YEARS} \\

\end{Large}
 
\begin{flushleft}


T. Sloan, (Dept of Physics, University of Lancaster)\footnote{Corresponding author: email t.sloan@lancaster.ac.uk}\\
A.W. Wolfendale, (Dept. of Physics, University of Durham) \\

\end{flushleft}
\end{center}


\begin{abstract}
\noindent
The Galactic cosmic ray (GCR) intensity has been postulated 
by others to vary 
cyclically with a peak to valley ratio of $\sim$3:1, as the Solar 
System moves from the Spiral Arm to the Inter-Arm regions of the 
Galaxy. These intensities have been correlated with global temperatures 
and used to support the hypothesis of GCR induced climate change. 
In this paper we show that the model used to deduce such a large 
ratio of Arm to Interarm GCR intensity requires unlikely values of 
some of the GCR parameters, particularly the diffusion length in the 
interstellar medium, if as seems likely to be the case, the diffusion 
is homogeneous. Comparison is made with the existing gamma 
ray astronomy data and this also indicates that the ratio is not 
large. The variation in the intensity is probably of order 10\,-\,20\% 
and should be no more than 30\% as the Solar System 
moves between these two regions, unless the conventional parameters 
of the GCR are incorrect. In addition we show that the 
variation of the GCR intensity, as the trajectory of the Solar System 
oscillates about the Galactic Plane, is too small to account 
for the extinctions of species as has been postulated unless, again, 
conventional assumptions about the GCR parameters are not correct.

\end{abstract}




 

\section{Introduction}

The rate of variation of the GCR intensity over the last 1000 million 
years (Ma) was investigated by Shaviv (Shaviv 2003). He deduced from a model 
that the GCR intensity varies by a factor $\sim$3 and cyclically with 
a period of 143 Ma as the Solar System moves from the Spiral Arm (SA) 
to the Inter-Arm (IA) region of the Milky Way Galaxy.  This variation was 
compared with proxies for the global temperature and a 
correlation observed (Shaviv and Veizer 2008). This correlation 
has been used as evidence favouring the contentious claim of a connection 
between GCR and climate change (Shaviv and Veizer 2008, Kirkby 2007, 
Kirkby 2012, Svensmark 2007). 

Subsequently, it was shown (Overholt et al., 2009) that the crossing 
of the spiral arms was in fact irregular and that the maximum and 
minimum GCR intensities deduced by Shaviv (Shaviv 2003) did not 
correlate in time with the Solar System crossings of SA and IA 
regions of the Galaxy. The intensity variation with time was 
studied from meteorites by Leveille et al 1999. From this study  
evidence was presented for an increase 
in the GCR rate during the last 10 Ma compared with the rate over 
the range 170-700 Ma (Leveille et al 1999). However, this result was 
not confirmed by Ammon et al (2009). They could not find evidence 
for a strongly varying GCR rate but their conclusion was based on only 
two meteorites. Wieler et al (2010) also could not find evidence for 
a strongly varying GCR rate from a study of iron meteorites 
although they did not report a statistical analysis. They used 
the same sample of the $\sim$80 meteorites reported by Vosage et al
(Vosage 1979 and Vosage 1984)    
as Shaviv (Shaviv 2003), each selecting different subsamples of 38 and 
50 meteorites, respectively.   
All these analyses are based on such small numbers of meteorites 
and the statistical precision of the data and the accuracy with which 
meteorite ages can be measured limit the conclusions which can  
be drawn from such studies. 


In this paper we show, using conventional assumptions of GCR parameters, 
that the difference in the GCR intensity as the Solar System moves from 
the SA to IA regions should be much smaller than that deduced by Shaviv. 
We use the data from gamma ray astronomy to confirm these findings. 
We go on to show that, unless GCR diffusion properties of the Galactic 
Interstellar medium are very different from their conventional form, 
GCR are unlikely to be associated with large scale climate changes such 
as the ice age epochs of the last billion years.    

\section{COSMIC RAY VARIATIONS OVER SPIRAL ARM/INTER-ARM TRANSITS.}

\subsection{The Spiral Geometry of the Galaxy.}
\label{describe}

                    The conventional picture of our Galaxy, 
the Milky Way, (and many other galaxies) is that it has 
spiral arms, these being regions where new star formation mainly takes 
place. In turn, short lived massive stars which later explode to form 
Type 2 Supernovae (SN), are mainly found in the Arms. Such SN remnants 
are generally thought to be responsible for the production of GCR 
with a differential injection energy spectrum falling as $E^{-n}$,   
with spectral exponent $n\simeq$2.15 up to energies $E\sim10^{15}$ eV. 
The current view is that the extra star formation is caused 
by the increased gas pressure in the Arms from the spiral density wave.  
In the present vicinity of the Solar System, the centre lines of the 
 SA are separated by about 1.7 kpc between the Local Arm and 
the neighbouring Sagittarius-Carina arm (Gies and Helsel, 2005 and  
Vall\'ee 2005).

 Separations between the SA nearer the Galactic Centre 
and in the far Outer 
Galaxy vary somewhat in the range 1.7 to 3 kpc. The variability arises 
from the inevitability of a non-perfect spiral wave, the gas density being
non-uniform in the pre-Galactic environment, together with tidal shear
from other galaxies. The adopted form of the Spiral Arms at present comes
from optical and radio measurements of stars and gas as a function of
Galactic longitude and latitude.

   Examining the distributions of the positions of SN remnants shows 
that they are roughly distributed about the the centre lines of the SA as 
Gaussian shapes with probable long tails. Such long tails will reinforce 
our conclusions of a small difference in the Galactic GCR intensity between 
the IA and SA. However, we make the conservative assumption of a Gaussian 
shape in what follows. The spatial distribution of Type 2 SN has been 
determined (Bartunov et al., 1994) to have a half-width at half 
maximum of $\sim$0.7 kpc along a Galactic radius, i.e. approximately 
a Gaussian with standard deviation of $\sim$0.6 kpc.

\subsection{Model predictions for the ratio of the IA and SA GCR intensities.}

\subsubsection{General Aspects}
\label{model}

                      To a first approximation one can assume that GCR are 
produced randomly in time and space in the SA but modulated by the radial 
distribution described in section \ref{describe}.  The GCR then 
diffuse with a spatially independent diffusion coefficient. Such 
a model is that used previously by us (eg Erlykin and Wolfendale, 2003), 
although without an SA/IA modulation. In that work a GCR scale height 
of 1 kpc was adopted. This is an important parameter in diffusion 
theory and is discussed in detail in section \ref{scht}. 

                      For a separation of the Arms (radially) of $d$, 
and a standard deviation equal to the scale height, $\sigma$, the GCR  
intensity, $I$, at the centre of one of several parallel equally spaced 
Arms will be, (adding the contributions from neighbouring arms):
\begin{equation}
  I(SA) = G(\sigma,0) + 2G(\sigma,d) + 2G(\sigma,2d) +...,
\label{eq1}
\end{equation}
where $G(\sigma,x) \propto \exp{-(x-\bar{x})^2/2\sigma^2}$ is the 
Gaussian function of $x$ about its mean $\bar{x }$. 

         Similarly, the GCR intensity at the centre of the IA will be:
\begin{equation}
I(IA)  =  2G(\sigma, 0.5 d) + 2G (\sigma, 1.5 d) + 2G( \sigma, 2.5d) +........
\label{eq3}
\end{equation}

                       Estimates of the differences in the 
GCR intensities, $I(IA)$ and 
$I(SA)$,  based on equations \ref{eq1} and \ref{eq3} are given 
for different values 
of $d$ and $\sigma$ in table 1. These are expressed as 
deficits, $\delta$, given by, 
\begin{equation}
\delta = 1 - \frac{I(IA)}{I(SA)}.
\label{delta}
\end{equation}

\begin{center} 
{\bf Table 1}

\noindent
Calculated deficits for different values of scale height, $\sigma$  
(standard deviation of particle diffusion) and $d$ the radial 
separation of the Spiral Arms.

\begin{tabular}{|c|c|c|}\hline
 $\sigma$(kpc) & $d$(kpc) & deficit($\delta$\%) \\
\hline
        1      &   2      &   3  \\
        1      &   3      &   36 \\
        2      &   2      &   0  \\
        2      &   3      &   0.1 \\
\hline
\end{tabular}
\end{center} 

The rather small calculated deficits occur because the inter-arm separations 
are of similar magnitudes to the scale heights, $\sigma$ 
(see section \ref{scht}). The conclusion from table 1 is that only for 
SA separated by more than 
2.5 kpc would we expect a deficit of more than 20\%.  A detailed 
comparison with the experimental data will be given later.

                           Keeping with our simple model (of constant 
diffusion coefficient, etc) attention can be drawn to the calculations 
of Erlykin and Wolfendale (2003) in which different modes of propagation 
were considered and expected proton spectra were estimated for randomly 
distributed SN in space and time, the GCR being assumed to come from the 
subsequent supernova remnants. Spiral Arm features were not considered but 
the spread of predicted spectra would correspond to different local 
locations of the SN. At GeV energies   
the range of predicted intensities was $\pm$20\% for normal 
diffusion and $\pm$30\% for 'anomalous' diffusion. The lower extreme values 
will correspond roughly to the Spiral Arm modulation situation so, 
again, a deficit of about 20\% is indicated. This value is of the same 
order as those indicated in table 1.

\subsubsection{The Cosmic Ray Scale Height}
\label{scht}

The scale height, $\sigma$, is defined as the distance  
from the median Galactic Plane at which the GCR 
intensity falls to a fraction e$^{-1/2}$ of its 
mid-Plane magnitude. The value is a convolution of the standard 
deviation of the source distribution ($\sim$0.6 kpc for super-novae, 
see section \ref{describe}) and the diffusion length of the produced 
GCR from the sources. It is appreciated that the distribution may not be
accurately Gaussian (see section \ref{describe} and later) but it is 
usually assumed to be so. Many
analyses give values of $\sigma$ from 1 to 2 kpc but others give larger 
values (e.g. Moskalenko et al., 2004 give 4-6 kpc).
It can be seen from equations \ref{eq1} and \ref{eq3} 
that the deficits decrease rapidly as the scale height increases. 

     In view of the standard deviation of the SN distribution being 
0.6 kpc (see
section \ref{describe}) it would be impossible for $\sigma$ to be 
less than this.
Much higher values than 1 kpc are not ruled out, however: explanations 
for the small 'Galactic gradient' of the GCR intensity, particularly in 
the Outer Galaxy, i.e. for Galactocentric distances greater than that for 
the Sun at radius 8.5 kpc,  include the possibility of a big scale height 
(eg Erlykin and Wolfendale, 2011, and earlier references therein).  
Indeed, Strong et al., (2004) suggest a value as high as 20
kpc. However, this could be due to the existence of a 2-component Halo 
with the Outer, low density region having the very large scale height. 
The GCR intensity distribution above and below the Plane could then still 
be close to that for $\sigma$ = 1 kpc.

Hunter et al (1997) fitted a comprehensive GCR propagation model to the 
EGRET data on the measured cosmic ray gamma ray intensities.  The fit 
gave a GCR scale height of 1.8 kpc. A useful further estimate of the 
scale height at the Galactic radius 
of the Solar System comes from radio astronomy (Erlykin and Wolfendale 2003).  Here, Beck (2009) gives $\sigma$ = 1.7 kpc, assuming that there is 
equipartition in energy between GCR and magnetic fields and a 
spatially constant GCR proton to electron ratio.

     We conclude that, at the Galactic radius of the 
Solar System, values of $\sigma$ outside the range 1 to 2 kpc are unlikely. 

\subsection{Implied deficits from Gamma Ray Astronomy}

\subsubsection{Principle of the method.}

High energy gamma rays are mainly produced by the interactions of primary 
GCR with the gas in the interstellar medium. Most gamma rays of energy 
above 0.3 GeV are produced by GCR primaries of energies of order several 
GeV. Specifically, Fathoohi et al.(1995) quote mean proton energies of 
2.6, 9.0,40 and 200 GeV to produce gamma rays of mean energy 0.3, 1.0, 
3.0 and 10.0 GeV, respectively. Thus, gamma rays of energy above 1 GeV 
come mainly from protons of energy above 9 
GeV (average $\sim$20 GeV), i.e. the energies responsible for much of 
the ionization in the Earth's atmosphere. Since the inter-stellar 
medium is transparent to gamma rays of this energy their  
intensity depends on the primary GCR intensity.  
 The gas column densities are known with reasonable 
accuracy and the gamma ray intensity along a line of sight 
after correction for the gas densities then gives an  
estimate of the primary GCR intensity.

        Inspection of maps of the Galaxy, eg as given 
by Gies and Helsel (2005) and by Vall\'ee (2005), shows that 
certain lines drawn from 
the Sun at particular longitudes pass largely through IA regions.  
The two most suitable longitudes are 
60$^\circ$ and 270$^\circ$. At these longitudes there is good SA, IA 
contrast. The measured gamma ray intensities in these directions will 
be compared with those in directions pointing towards the nearby SA 
to estimate the deficits.  

\subsubsection{Results from COS B measurements.}

 The search for SA, IA differences is not a new one.
The COS B satellite (Bignami 1975, Bennett et al., 1976) gave early 
relevant gamma ray measurements and it is appropriate to mention 
the work here.

     We (Rogers et al., 1988) and Bloemen et al.,(1989) presented 
evidence for the spectral 
shape of GCR depending on Galactic latitude and SA, IA intensity 
differences. Our own work gave a difference of spectral exponent between 
SA and IA of 0.4$\pm$0.2 for the Orion Arm and its neighbouring IA. Thus an 
increasing deficit with increasing gamma ray energy was indicated. 
      An analysis of the IA, SA 
contrast in CR intensity  was also given by Van der Walt and Wolfendale 
(1988). These workers found values for the deficit in the range 10 to 35\% 
for gamma rays of relevance to the work described here.

It should be remarked, however, that these results can only be regarded 
as indicative of an IA, SA spectral difference for two reasons:- 

a) The latitude range is too high for our present purpose; Rogers et al 
(1988) use a mean latitude of 10$^\circ$ (compared with 0$^\circ$ in the 
next section) and here the inverse Compton contribution will also cause 
difficulties. 

b) The later data from EGRET (section \ref{EGRET}) are more accurate.

\subsubsection{Results from EGRET measurements.}
\label{EGRET}
      The EGRET data have been analysed by Hunter et al., (1997) 
who gave useful longitudinal distributions of gamma ray 
intensities for a range of gamma ray energies and latitudes. 
These include measurements in the Galactic 
Plane with -2 deg$< b <$ 2 deg, where $b$ is the Galactic latitude. 
Their distributions of gamma ray intensity as a function of Galactic 
longitude in this plane are used here to estimate the deficits. 

To achieve this we take the difference in gamma ray intensity 
in the Galactic Plane at a  
longitude pointing towards a region dominated by an IA from a  
region pointing to a nearby SA. For the 60$^\circ$ IA region 
comparison is made with the region centred on 45$^\circ$ 
and the 270$^\circ$ IA region is compared with the 
285$^\circ$ direction, the 45$^\circ$ and 285$^\circ$ regions 
each pointing towards the Sagittarius-Carina (S-C) Galactic Arm.  

      Table 2 gives the deficits estimated from the EGRET data for the 
two highest energy bands: 0.3 to 1 GeV and $>$ 1 GeV.
The deficits are estimated by taking the difference of 
the EGRET intensities divided by the column densities of gas 
(from Burton (1976) and Levine et al., (2006)). A correction has been 
applied for the fact that at each longitude there is a contribution 
from both SA and IA regions (method a).   A second method (method b) 
uses the fit given in Hunter et al (1997).  In this fit it was found 
that over much of the relevant Galaxy a model in which 
there was a correlation of GCR intensity with total gas density (with a 
spatial smoothing)  gave a good fit. Thus the fitted gamma ray intensity 
profile 
is used to estimate the deficits by subtracting the fitted 
intensities in each direction. 

\begin{center} 
  {\bf Table 2.}           

\noindent
Spiral Inter-Arm Directions from the Sun. \\
Key - Arms; S-C  Sagittarius-Carina; $L^\circ$ is the Galactic longitude 
(90-270$^\circ$ looking outwards from Galactic Centre); 
$d$ is the radial separation 
of the centres of SA; the estimated deficit is $\delta$ 
(see equation \ref{delta} for definition).  The uncertainties in 
each case are about 10\%.  The values given refer to 
method (a)  and those in brackets to method (b) (see text).

\begin{tabular}{|c|c|c|c|c|}\hline
$L^\circ$ & Adjacent Arms & $d$ (kpc) &    $\delta$(\%) (0.3-1 GeV) 
& $\delta$(\%) ($>$ 1 GeV.) \\
\hline
 60         &  Local, S-C and Perseus & 2.7 &    25(18) &  20(18)     \\
 270        &  S-C and Perseus & 3.0 &  29(26) &  25(30)      \\

\hline
\end{tabular}
\end{center} 

It should be noted that the structures expected as the longitude changes 
from an IA to an SA direction are not convincingly present in 
the EGRET data. The data show a smooth downward trend from the Galactic 
Centre together with diffuse structures left over from point sources 
from which the direct radiation has been removed. Hence the majority of 
the deficit shown in Table 2 is due to the smooth decrease of intensity 
with longitude away from the Galactic Centre.     
The deficits shown in Table 2 are from the values read off the data 
as if the expected structures were present. Hence these values are  
upper limits on the deficits.

                         That the results are 'reasonable' can be seen 
by examining the plot from Hunter et al. (1997) of the 'CR enhancement 
factor' along the near-circular track of the Solar System along its path 
round the Galaxy. Only two regions of excess were found 
and these gave an effective deficit of 30\%. However, 
these regions of excess do not correlate well with the known location of 
the Spiral Arms (Gies and Helsel 2005, Vall\'ee 2005). 

\subsubsection{Results from FERMI-LAT measurements.}

        The most recent gamma ray results come from the FERMI-LAT 
observatory (Mizuno et al., 2011). These workers have examined the local 
region of the Outer Galaxy: specifically the Local Arm and the Perseus 
Arm regions as well as the IA region between the two. For gamma ray 
energies of 
greater than 0.3 GeV they found similar intensities from the Perseus 
arm and the IA region, implying a deficit of around zero. They found  
15\% less emission from the IA region than that in the Local Arm, 
implying a deficit of 15\%. Hence, these values indicate deficits 
due to IA, SA differences of less than or order 15\%.    

\subsubsection{Other Galaxies}

Observations of the structure of our Galaxy from a position within it 
(i.e. the Earth) are difficult and, in principle, recourse can be made 
to other galaxies. Some galaxies, viewed end-on, have impressive haloes 
but the measurements yield information about electrons, and magnetic 
fields and electrons, of course, suffer extra energy losses so that their 
scale heights will be smaller than those for protons. 

Of greater interest is the recent measurements of the Large Magellanic 
Cloud (LMC) by Abdo et al.(2010) using Fermi-LAT. These workers found what 
appeared to be a scale height of 0.2 kpc for cosmic rays responsible 
for gamma rays in the 0.2-20 GeV energy range. The primary cosmic rays 
are plausibly (authors' term) the origin of the gamma rays by way of 
interactions with the inter-stellar medium and the radiation fields. 
This implies both cosmic ray protons and electrons. Such a small scale 
height, if applied to our own Galaxy, would imply a very large deficit  
according to equations \ref{eq1} and \ref{eq3} with almost zero GCR 
intensity when the Solar System is outside the SA. We examine this 
possibility. 

Firstly, the phenomenon is restricted to the environs of 30 Doradus, 
the well known, ultra-active star forming region. Now many of the 
properties of the LMC are very different from those in the Galaxy (see 
Chi and Wolfendale, 1993) and, as these workers showed, the ambient 
cosmic ray flux is down by a factor 5-10 compared with the local flux. 
The recent Fermi-LAT measurements have refined this to a factor $\sim$2-4. 
Most importantly, Abdo et al (2010) found that the gamma ray emission 
from the LMC shows little correlation with the gas density. Concerning 
30 Doradus, it is likely that strong stellar winds play a role in 
driving cosmic rays out of this small galaxy. 

That there is nothing equivalent in our Galaxy comes from comparing 
the Galactic longitude distribution of gamma rays with that of molecular 
hydrogen - which is an indicator of cosmic ray production. Strongly 
active Galactic regions such as those associated with the Cygnus 
complex do not show the 30 Doradus phenomenon. 

We consider that the information about the scale heights from other 
galaxies is not useful.

\subsection{Discussion of the GCR Arm, Inter-Arm intensities.}

  The model adopted (see Table 1) gave predicted 
values of $\delta$ in the range 0 to 36\% for $\sigma$ from 1 to 2 kpc 
and for Arm separations of 2 to 3 kpc. The values 
are, necessarily approximate, for a number of reasons as well as 
the inaccurately known parameters:

             a. Problems with likely differences in diffusion 
coefficients in the Arm and Inter-Arm regions.

             b. Stochastic differences due to the actual distribution of 
relevant SN (and SN remnants) as distinct from their average values.

             c. Possible effects of Galactic Winds.

                                        Turning to the experimental 
observations, the derivation of the values of $\delta$ is necessarily 
imprecise. Foreground 
contributions are not easy to estimate. Nevertheless, a comparison of 
'observed' and 'expected' is made.

Taking the data in Table 2 at face value and averaging we obtain an 
observed $\delta$ = 23 $\pm$5~\%.  Using the model, with 
$\sigma$ = 1 kpc and the mean value of $d$ we predict 
$\delta$ = 25 $\pm$ 10 \%. There is thus reasonable  
agreement between the model and the data. Note that the model will 
give a much lower value of the deficit if the GCR scale height is 
closer to the value of 1.8 kpc from the fit to the EGRET data. 
This is not incompatible with the data in Table 2 which should 
be looked on as an upper limit rather than a measurement.    

   We conclude that the deficit in Inter-Arm versus Arm intensities over the 
Galactic circuit of the Solar System is most unlikely to have exceeded 30\%    
at cosmic ray energies of several GeV, the energy producing the majority 
of the ionization in the Earth's atmosphere. 

\section{Influence of changes in the IA/SA GCR intensity 
on the climate}

We have shown above that the changes in the GCR intensity as 
the Solar System moves from the IA to the SA of our Galaxy are  
of the order of 20\%. These match roughly the changes in the 
intensity observed during the 20th century due to centennial effects 
and due to the 11 year solar modulation. It has been shown that 
changes at this level cause at most a 0.07$^\circ$C change in the 
present day mean global temperature (Erlykin et al., 2009b). 
Therefore, the changes in global temperature due to changes in the 
GCR intensity as we move from 
the IA to the SA are likely to be of the same magnitude. 
We conclude therefore that such changes cannot produce the large 
changes at the Ice Age epochs of the past 10$^9$ years as postulated 
by Shaviv and Veizer (2008).  

\section{Cosmic Ray changes due to vertical
oscillations of the Solar Cycle and species extinctions.}

          Just as the Solar System 
passes through successive SA, so it oscillates about the Galactic  
Plane. The period is about 64 Ma (Gies and Helsel, 2005) and the 
amplitude is $\sim$70 pc (Thaddeus and Chanan, 1985). 
Interestingly, 
there is a cycle of 'fossil diversity' with a period of 62 $\pm$ 3 Ma 
(Rohde and Muller, 2005 and Melott and Bambach, 2011). Connecting the two 
observations, Medvedev and Melott (2007) 
have proposed that the changes in the GCR intensity due to these 
oscillations was responsible for the extinctions which produced 
the fossil diversity. Hence, 
again, there is a model to be tested.

       An immediate problem for the model is that with any reasonable GCR 
scale height (say 1 kpc, or above) the fall in GCR intensity at the 
extremes of the (vertical) oscillations will be negligible. As 
implied by section \ref{scht},  a scale height of the necessary 100 pc  
is not possible. Therefore, the changes in the GCR intensity due to 
such vertical oscillations will be too small to produce species 
extinctions. Hence it seems most unlikely that they will produce the 
fossil diversity as postulated by Medvedev and Melott (2007). 

      A further point is that  
there will be two Galactic Plane crossings per cycle and thus an 
expected fossil diversity period of 32 Ma rather than 64 Ma.
Medvedev and Melott (2007), endeavour to circumvent this problem 
by assuming an anisotropy in the GCR intensity due to extragalactic 
effects. However, such effects appear unlikely in view 
of a near-symmetry in Galactic latitude of the non-thermal 
radio signal (Broadbent et al., 1989). 

\section{Conclusions}

The Shaviv model of the difference in GCR rate in the IA and SA 
regions of the Galaxy gives large values for the change in the 
cosmic ray intensity in passing from the Spiral Arm of the Galaxy to the 
Inter-Arm region. The calculations presented here, using conventional 
values for GCR diffusion properties, give such changes at the level of
10 to 20\% rather than the factor $\sim$3 expected from his 
model. The data from gamma ray astronomy are compatible with the 
smaller changes presented here and are incompatible with the large 
changes proposed in Shaviv's model. 
Furthermore, Wieler et al. (2011) showed from meteorite data 
that in the last 10 Ma the GCR intensity has not 
varied by more than 10\%. Now, some 10 Ma ago the Solar System was 
well into the IA region compared to its position in the Local Arm now 
so that the probability of the local SA, IA 
deficit being as high as or greater than 20\% must be small. 

It has already been shown by Overholt et al (2009)
that the peaks and troughs in the Shaviv distribution do 
not correspond to crossings of the SA in the Galaxy. Here we 
show that the estimated intensity variations from the 
Shaviv distribution are also unrealistic, if we use conventional 
assumptions of the GCR parameters.   

We conclude therefore that the use of this model to claim that 
there is palaeontological evidence for a connection between cosmic 
rays and the climate is unjustified, unless some of the conventional 
assumptions of GCR parameters are wrong.   
Added to our earlier analysis of the near contemporary GCR and Global 
temperature measurements (Erlykin et al 2009a and 2009b)   
which showed no evidence for a GCR-climate link we conclude that there 
is no hard evidence for such a link. 

An explanation of the reason for the initiation of the Ice Age 
epochs of the past 10$^9$ years remains to be found. Such 
initiation may possibly have an astronomical cause  
by way of the effect of GCR of PeV energies 
on the electrical conditions of the atmosphere. The periods 
of increase here are of order 20 ka occurring every Ma or so. 
The 20 ka period arises  
from the rapid diffusion of the PeV particles which are deemed 
responsible. The fact referred to 
by Erlykin and Wolfendale (2001) is that proximity to Supernova 
remnants in the Solar System's passage through the Galaxy causes 
increases by several orders of magnitude in the intensity of 
terrestrial 
GCRs at PeV energies (as distinct from a few percent in the GeV 
energy range). This might 
have relevance, the multiplying factor coming by way of effects 
on the electrical conditions of the atmosphere. 

\section{Acknowledgements}

We thank B. Bottke, A.D. Erlykin and J. Kirkby for 
stimulating and useful discussions. We acknowledge the support 
from our home institutions 
and we thank the John Taylor Foundation for financial support.

\end{document}